\documentclass[11pt]{article}
\usepackage[margin=1in]{geometry}
\usepackage{graphicx}
\usepackage{hyperref}
\usepackage{booktabs}
\usepackage{amsmath}
\usepackage{listings}
\usepackage{xcolor}

\title{SMS Opt-In/Opt-Out Consent Record Architecture in Enterprise CRM Systems: Compliance Patterns for Multi-Tenant Managed Packages}
\author{Devam Gupta \\ Staff Engineer, Twilio \\ Technical Architect, Twilio for Salesforce}
\date{July 2026}

\begin{document}
\maketitle

\begin{abstract}
Regulatory frameworks such as the Telephone Consumer Protection Act (TCPA) impose strict consent requirements on enterprise messaging systems: organizations must obtain and record explicit opt-in consent before sending SMS communications, and must immediately honor opt-out requests. While CRM platforms provide rich contact and lead data models, they do not natively model consent state as a first-class record type. This gap becomes architecturally significant in multi-tenant managed packages distributed via enterprise application marketplaces, where the package cannot assume or modify the installing organization's schema. This paper presents a production consent record architecture, generalized from patterns implemented within a CRM-native messaging managed package serving independent enterprise organizations spanning healthcare, financial services, and sales operations. We describe the data model design, keyword-based consent capture, a hash-based uniqueness strategy for deduplication, suppression enforcement at message send time, and the multi-tenant constraints that shaped these design decisions.
\end{abstract}


\section{Introduction}

Consent management in enterprise messaging presents a layered problem. At the regulatory layer, TCPA requires documented, per-contact opt-in before commercial SMS, with immediate opt-out processing upon receipt of standard keywords (STOP, QUIT, CANCEL, UNSUBSCRIBE) and re-opt-in on receipt of START or UNSTOP. At the data layer, CRM systems such as Salesforce model contact records extensively but do not include a native consent state object -- opt-in status is typically tracked via checkbox fields on Contact or Lead records, which does not scale to multi-keyword, multi-channel consent architectures. At the integration layer, a managed package installed across many independent organizations cannot depend on custom fields added by individual organizations, cannot modify core CRM schema, and must operate within platform resource limits that constrain query patterns at bulk message send time.

These constraints together motivate the design described in this paper: a dedicated consent record object with a polymorphic relationship to both Contact and Lead standard objects, keyword-scoped consent tracking, a deterministic hash-based unique identifier for deduplication, and a suppression check integrated into the bulk message send pipeline at the service layer rather than the trigger layer.

This paper is a companion to two earlier papers on the same class of production architecture: a treatment of the overall CRM-native messaging data model and send/receive pipeline~\cite{gupta2026arch}, and a treatment of webhook-based status reconciliation for outbound message delivery~\cite{gupta2026webhook}. Where those papers describe the system's general data model and its status-tracking reliability guarantees, this paper focuses specifically on the consent-and-compliance problem: how such an architecture can enforce regulatory opt-in/opt-out requirements correctly and efficiently at bulk scale, within the schema constraints imposed by managed-package distribution.

\section{Related Work}

Prior work on enterprise consent management has largely focused on regulatory compliance frameworks and legal interpretations of TCPA rather than technical implementation. Solove and Hartzog~\cite{solove2014} provide a comprehensive analysis of consent in digital privacy law but do not address the technical data-model constraints imposed by managed package distribution. Similarly, Cranor~\cite{cranor2012} examines consent in web contexts without consideration of CRM-native architecture.

In the CRM and enterprise integration space, the companion paper on CRM-native messaging architecture~\cite{gupta2026arch} describes the webhook decoupling and asynchronous send pipeline that forms the delivery layer on top of which the consent enforcement described here operates, and the companion paper on webhook reliability~\cite{gupta2026webhook} describes the two-path status verification mechanism that guarantees delivery-state consistency for the same class of system. The consent architecture described in this paper is designed to integrate with that delivery pipeline at the service layer.

Work on multi-tenant SaaS architecture~\cite{bezemer2010} addresses schema isolation and tenant-specific configuration, but does not address the specific constraint of managed-package schema immutability -- where the ISV cannot modify the installing tenant's schema and must operate entirely within the package's own custom object layer.

The hash-based external-ID pattern for idempotent upsert in distributed systems draws on established practices in event sourcing and CQRS architectures~\cite{young2010}, applied here to the specific context of a platform's native upsert mechanism and resource-limit constraints.

\section{Data Model Design}

\subsection{Dedicated Consent Object}

The consent architecture centers on a dedicated custom consent object rather than fields on the Contact or Lead standard objects. This decision is driven by two constraints specific to managed packages.

First, adding fields to standard objects (Contact, Lead) in a managed package installs those fields into every customer organization. Fields that may conflict with existing customer field names, or that may create confusion for customers not using keyword-based consent, should not be imposed on the standard object schema.

Second, a dedicated object supports multiple consent records per contact -- one per keyword -- which is required when a single contact opts in to multiple messaging programs (e.g., a promotional keyword and a transactional keyword) from the same organization.

Key conceptual fields on the consent object are summarized in Table~\ref{tab:fields}.

\begin{table}[h]
\centering
\caption{Key Fields on the Consent Record Object}
\label{tab:fields}
\begin{tabular}{p{4cm}p{9cm}}
\toprule
\textbf{Field (conceptual)} & \textbf{Purpose} \\
\midrule
Contact reference & Polymorphic lookup -- set when the consenting party is a Contact \\
Lead reference & Polymorphic lookup -- set when the consenting party is a Lead \\
Keyword reference & Which consent keyword triggered the opt-in \\
Active flag & Current consent state \\
Opt-in timestamp & When opt-in was recorded \\
Opt-out timestamp & When opt-out was recorded (if applicable) \\
Source message identifier & Provider message identifier of the opt-in message \\
Phone number (raw) & Raw phone number as received \\
Phone number (normalized) & Digits-only normalized form \\
Composite unique key & Hash of normalized phone number and keyword identifier \\
\bottomrule
\end{tabular}
\end{table}

\subsection{Polymorphic Contact/Lead Relationship}

Campaign members in a CRM can be either Contacts or Leads. The consent architecture mirrors this by maintaining separate lookup fields rather than a single polymorphic reference pattern, which simplifies query construction and avoids the additional type-dispatch logic that a unified polymorphic lookup would require. At query time, callers branch on which reference is populated:

\begin{lstlisting}
for (consentRecord in consentRecords):
    if consentRecord.contactRef is not null:
        contactIds.add(consentRecord.contactRef)
    elif consentRecord.leadRef is not null:
        leadIds.add(consentRecord.leadRef)
\end{lstlisting}

\subsection{Hash-Based Unique Identifier}

A core requirement is idempotent upsert: when an inbound SMS opt-in message arrives, the system must either create a new consent record or update an existing one without creating duplicates. CRM platforms support external-ID-based upsert operations that satisfy this requirement, but the external ID field must be deterministically computable from the inbound data alone.

The solution is a composite hash stored as the record's external identifier, constructed as:

\[
\text{UniqueId} = \text{normalize}(\text{phoneNumber}) \,\Vert\, \text{keywordId}
\]

where $\text{normalize}(\cdot)$ strips all non-digit characters to reduce E.164, locally-formatted, and unformatted phone numbers to a common digits-only representation, and $\text{keywordId}$ is the platform-assigned identifier of the keyword record. This produces a unique string per (phone number, keyword) pair within an organization that is stable across repeated inbound messages from the same number.

The composite key field is designated as an external ID, enabling the platform's native upsert mechanism, executed through the same centralized, permission-enforcing data-access layer described in~\cite{gupta2026arch}:

\begin{lstlisting}
dataAccessLayer.upsert(
    consentRecords,
    externalIdField = "UniqueId",
    enforcePermissions = true
)
\end{lstlisting}

Phone number normalization is computed at record write time (before insert and before update) via a trigger handler, ensuring that the stored normalized form is always consistent regardless of how the number was originally formatted in the inbound message.

\section{Keyword-Based Consent Capture}

\subsection{Keyword Object}

Consent keywords are modeled as a separate lookup object with a name field storing the keyword string (e.g., ``JOIN'', ``SUBSCRIBE'', a campaign-specific keyword). This decouples keyword management from the consent records themselves and allows an organization to manage multiple active keywords independently.

\subsection{Inbound Keyword Matching}

When an inbound SMS arrives via the webhook path described in~\cite{gupta2026arch}, the message body is inspected against the set of configured keywords for the organization. The keyword matching is case-insensitive and applied against normalized message-body content. If a match is found, a consent record is upserted with its active flag set to true.

For opt-out keywords (STOP, QUIT, CANCEL, UNSUBSCRIBE, and variants), the matching logic deactivates all active consent records associated with the sending phone number, regardless of keyword:

\begin{lstlisting}
function deactivateConsentByPhoneNumber(phoneNumbers):
    records = consentSelector.getActiveByPhoneNumber(phoneNumbers)
    for record in records:
        record.active = false
    dataAccessLayer.update(records, enforcePermissions = true)
\end{lstlisting}

This implements the regulatory requirement that a STOP to any number in the messaging program must suppress all messages from that program to that number, not just the specific keyword channel.

\subsection{Re-Opt-In Logic}

Re-opt-in after a prior opt-out requires distinguishing a new opt-in event from a duplicate of the original. The upsert logic checks three conditions before reactivating a record:

\begin{lstlisting}
if (existing.active == false
    and incoming.optInTimestamp > existing.optOutTimestamp
    and incoming.sourceMessageId != existing.sourceMessageId):
    incoming.active = true
\end{lstlisting}

The timestamp comparison ensures the new opt-in event is more recent than the opt-out. The source-message-identifier comparison ensures the inbound message triggering the re-opt-in is distinct from any previously processed message -- preventing a webhook replay or retry from incorrectly reactivating a record. This idempotent-convergence approach parallels the state-machine convergence guarantee used for delivery-status reconciliation in~\cite{gupta2026webhook}.

\section{Suppression Enforcement at Send Time}

\subsection{Service Layer vs. Trigger Layer}

Consent suppression -- the act of preventing a message from being sent to an opted-out contact -- is enforced in the messaging service layer, not in a trigger. This placement is deliberate. Trigger-layer suppression would require every insert of a message record to query the consent object, introducing a query per DML operation and creating resource-limit pressure at bulk send volumes (hundreds of records per transaction). Service-layer enforcement queries opt-in state once per batch before any DML occurs, making the check scalable to bulk message sends of arbitrary size.

\subsection{Campaign Member Opt-In Check}

For bulk campaign messages, the suppression check operates on the full set of campaign-member records before any messages are created:

\begin{lstlisting}
function getConsentedCampaignMembers(members):
    hashToMember = {}
    phoneNumbers = set()
    validMembers = []

    for member in members:
        if member.campaign.optInKeyword is not null:
            hash = buildUniqueHash(
                normalize(member.mobilePhone),
                member.campaign.optInKeyword)
            hashToMember[hash] = member
        else:
            validMembers.append(member)  # no keyword => no consent gate

    activeConsents = consentSelector.getActiveByHashSet(
        hashToMember.keys())

    for consent in activeConsents:
        if consent.uniqueId in hashToMember:
            validMembers.append(hashToMember[consent.uniqueId])

    return validMembers
\end{lstlisting}

The hash-keyed lookup ensures a single query retrieves all relevant consent records for the entire batch, with $O(n)$ map lookups to filter the campaign member list.

\section{Multi-Tenant Constraints}

\subsection{No Assumptions on Installing Org Schema}

The consent architecture is entirely self-contained within the managed package's custom object layer. No queries touch standard object fields that may be customized or restricted in individual customer organizations. Phone number fields are read from the standard mobile-phone fields present on Contact and Lead in all Salesforce orgs, with normalization applied in the package layer to handle formatting variations.

\subsection{Resource-Limit Architecture}

Managed multi-tenant platforms typically enforce a bounded number of database queries per transaction~\cite{sfgovlimits2024}. A bulk message send for a large campaign executes within a single transaction context managed by the platform's batch-processing framework. The consent check is structured as a single query against the consent object using a membership clause on the pre-computed hash set, regardless of campaign size:

\begin{lstlisting}
query += " WHERE active = true AND uniqueId IN :hashSet"
\end{lstlisting}

This ensures the consent check consumes exactly one query regardless of campaign-member count, leaving resource-limit headroom for the message-creation DML and any downstream operations.

\subsection{Async Deactivation}

Opt-out deactivation is executed asynchronously when invoked from synchronous trigger context, to avoid consuming DML statements in the trigger execution path:

\begin{lstlisting}
function deactivateConsentAsync(phoneNumbers):
    count = consentSelector.countActiveByPhoneNumber(phoneNumbers)
    if count > 0:
        if isBatchContext() or isAsyncContext():
            deactivateConsentByPhoneNumber(phoneNumbers)
        else:
            enqueueAsyncDeactivation(phoneNumbers)
\end{lstlisting}

The context check prevents the nested-async exception that many platforms raise when an asynchronous unit of work is enqueued from within an already-asynchronous execution context.

\section{Campaign Deliverability Integration}

Consent-state changes propagate to campaign deliverability tracking through the trigger domain handler. When a consent record's active flag changes, the system identifies all campaigns that include the affected contact or lead and marks those campaigns' deliverability status as stale, signaling that the deliverability count needs recalculation before the next send:

\begin{lstlisting}
for campaign in affectedCampaigns:
    campaign.deliverabilityStatus = "stale"
dataAccessLayer.update(affectedCampaigns, enforcePermissions = true)
\end{lstlisting}

This lazy-invalidation pattern avoids recalculating deliverability counts on every opt-out event -- counts are recomputed only when a send is actually initiated, at which point the stale flag triggers a fresh consent query.

\section{Discussion}

The architecture described here reflects constraints specific to managed-package distribution that are absent in single-tenant implementations. A single-tenant developer can add checkbox fields to a Contact object, use declarative automation to set suppression flags, and rely on org-specific automation. A managed package must treat each installing organization as a black box: the schema outside the package boundary is unknown, the permission configuration is unknown, and the data volume is unknown.

The hash-based external-ID pattern is the most consequential design decision in this architecture. It enables idempotent upsert without querying for existing records first, which would require two DML operations per consent event instead of one. At scale -- processing many inbound opt-in messages per hour across a large number of tenants -- this reduction matters for both performance and resource-limit consumption.

The separation of suppression enforcement into the service layer, rather than the trigger layer, reflects a general principle applicable beyond consent management: validation logic that must scale to bulk operations should not be placed in per-record trigger handlers, which accumulate query and DML usage proportionally with record count.

Read together with~\cite{gupta2026arch} and~\cite{gupta2026webhook}, this paper completes a three-part account of a class of production architecture: the general CRM-native data model and send/receive pipeline, the status-reconciliation mechanism that keeps delivery state consistent under failure, and the consent-and-compliance data model that governs which messages are permitted to be sent in the first place.

\section{Conclusion}

This paper has presented a production consent record architecture for enterprise CRM-native messaging, generalized from patterns implemented in a managed package deployed across independent enterprise organizations. The core contributions are: a dedicated consent object design that avoids standard-object modification constraints, a hash-based external-ID strategy for idempotent upsert across a distributed multi-tenant deployment, keyword-scoped consent with cross-keyword opt-out enforcement, and service-layer suppression enforcement designed for resource-limit compliance at bulk-send scale. These patterns address constraints specific to managed-package distribution and are applicable to any CRM-native messaging integration subject to consent regulatory requirements.

\end{document}